\documentclass[12pt,emtex]{article}
\usepackage{amsfonts}
\usepackage{amssymb}
\usepackage{amscd}
\usepackage{amsmath}

\newcommand{\BEQ}{\begin{equation}}
\newcommand{\EEQ}{\end{equation}}
\def\bea{\begin{eqnarray}}
\def\eea{\end{eqnarray}}
\def\nn{\nonumber}

\newtheorem{Th}{Theorem}
\newtheorem{lem}{Lemma}
\newtheorem{Lem}{Lemma}

\newtheorem{Rem}{Remark}

\def\bea{\begin{eqnarray}}
\def\eea{\end{eqnarray}}

\def\bes{\begin{equation*} \begin{split}}
\def\ees{\end{split} \end{equation*}}

\def\O{{\cal O}}

\def\d{\displaystyle}
\def\h{{\mathsf h}}

\def\one#1{#1^{\raise5pt\hbox{$\scriptstyle\!\!\!\!1$}}\,{}}
\def\two#1{#1^{\raise5pt\hbox{$\scriptstyle\!\!\!\!2$}}\,{}}

\def\t{\tau}
\def\g{\mathfrak{gl}_n}
\def\e{\widetilde{e}}

\def\p{\partial_u}

\begin{document}

\begin{titlepage}
\hfill ITEP-TH-14/04 \vskip 2.5cm

\centerline{\LARGE \bf
Quantization of the Gaudin System
}

\vskip 1.0cm \centerline{
D. Talalaev \footnote{E-mail: talalaev@aport.ru}
\footnote{This work has been partially supported by the RFBR grant
04-01-00702}}

\centerline{\sf Institute for Theoretical and Experimental Physics
\footnote{ITEP, 25 B. Cheremushkinskaya, Moscow, 117259, Russia.}}

\vskip 2.0cm

\begin{abstract}
In this article we exploit the known commutative family in $Y(\g)$ - the
Bethe subalgebra - and its special limit to
construct quantization of the Gaudin integrable system. We give
explicit expressions for quantum hamiltonians $QI_k(u),~ k=1,\ldots, n.$ At
small order $k=1,\ldots,3$  they coincide with the quasiclassic ones, even
in the case $k=4$ we obtain quantum correction.
\end{abstract}

\vskip 1.0cm

\tableofcontents

\end{titlepage}


\section{Introduction}
\paragraph{Quantization.}
The quantization problem has different aspects and different approaches
therein. The quantization consists in constructing a deformation of a
Poisson algebra of functions on a phase manifold and an appropriate
representation. We are interested in the first part of quantization problem
in a specific case of integrable systems. An integrable system is a
hamiltonian system on a Poisson manifold $(M^{2n},\pi)$ where $\pi$ is a Poisson
bivector, given by a hamiltonian $H$ such that there exists a family of
algebraically independent functions $I_1,\ldots,I_n$ in involution
$\{I_i,I_j\}=0$ and $H=I_1.$ In this language an integrable system is a
subalgebra $\Im\subset C(M)$ which is also commutative subject to the
Poisson bracket and is generated by the appropriate number of generators.
This property of being appropriate is even more ambiguous at the quantum
invariant level, so we do not discuss it here.

Traditionally the quantization of a Poisson algebra on $C(M)$ is an associative algebra
$A(M)$ isomorphic as a linear space to the space of formal series $C(M)[\hbar]$
with a fixed isomorphism of linear spaces $$q:C(M)[\hbar]\rightarrow A(M)$$
called a quantization correspondence, such that
$$q(f)*q(g)=q(fg)+O(\hbar)\quad\mbox{and}\quad
q(f)*q(g)-q(g)*q(f)=\hbar q(\{f,g\})+O(\hbar^2)$$
where $f,g$ are functions on $M,$ $*$ is the multiplication on $A(M)$ and
$\{f,g\}$ is the Poisson bracket. The quantization of an integrable system
in this context is such a quantization of a Poisson algebra that $q(\Im)$ is
a commutative subalgebra (not only at the first order in~$\hbar$) in $A(M).$
The importance of this condition is obvious, as like as in the classical
case the dynamics can be restricted to the common level of integrals, the
quantum problem of finding the eigenvectors can be restricted to the common
eigenspace of quantum integrals.

The main example of this paper is the quantization of the space $g^*$ for a
Lie algebra $g=\g$ with the Kirillov-Kostant Poisson bracket. The
space of functions in this case is just the  symmetric algebra
$C(M)=S^*(g).$ The quantum algebra is the universal enveloping algebra
$U(g)$ deformed in a standard manner: one takes new generators
$\widetilde{e}_{ij}=\hbar e_{ij}$ than the defining relations in $U(g)[\hbar]$ became
$$[\e_{ij},\e_{kl}]=\hbar(\delta_{jk}\e_{il}-\delta_{li}\e_{kj}).$$
The quantization morphism on homogeneous polynomials is given by
fixing some normal ordering. The classical limit consists in tending
$\hbar\mapsto 0.$

The same procedure can be realized without thinking of formal parameter due to the
structure of filtered algebra on $U(\g).$ Indeed, the quantization can be
considered as a map from $S^*(\g)$ to just $U(\g)$ given by the same normal
ordering. Let us note that the choice of such an ordering
does not affect the result of passing to the associated graded algebra
$Gr(U(\g))=S^*(\g).$ Hence one can consider the factorization morphism
$\rho:U(\g)\rightarrow Gr(U(\g))$ as a classical limit. Indeed, the
composition of quantization and factorization is an identity
on homogeneous element of the symmetric algebra $S^*(\g).$

\paragraph{Gaudin system.}
Let us consider the direct sum of $k$ coadjoint $\g$-orbits  $\oplus_i \O_i$
with an induced Poisson (symplectic) structure and the diagonal symplectic
$GL_n$-action. An element of this space can be represented by the Lax
operator
\begin{equation} \label{Lax}
L(u)=\sum_i \frac {\Phi_i}{u-z_i}
\end{equation}
with $\Phi_i\in\O_i.$ Fixing a
basis one can think of $\Phi_i$ as a matrix valued function on $\O_i$ with
matrix elements $f_{kl}$ - generators of $\g$ considered as linear functions
on the $i$-th orbit.
Then the functions $$I_i(u)=TrL^i(u)\quad\mbox{commute}\quad \{I_i(u),I_j(v)\}=0$$
and are invariant under the diagonal $GL_n$-action. The images of these functions on
the symplectic reduction space $M=\oplus_i \O_i//GL_n$ give the integrable
system. The proof of the integrability can be found for example in
\cite{CT1} and is based on the $r$-matrix representation of the Poisson
bracket:
$$\{L(u)\otimes L(v)\}=[r(u-v),L(u)\otimes 1+1\otimes
L(v)]\quad\mbox{with}\quad r(u)=\frac P u$$ where $P$ is a permutation
operator in the tensor product.
There exist several partial responses to the quantization problem of this
integrable system. For example in the $\mathfrak{gl}_2$ case just the same
expressions $$I_2(u)=Tr L^2(u)$$ with ${f}_{kl}$ changed to the
$U(\mathfrak{gl_2})$ generators $e_{kl}$ give the commutative family of elements
in $U(\mathfrak{gl_2})^{\otimes k}.$
The attempt to generalize this result  to $\g$ in \cite{CRT} was
only partially successful. There was obtained that quantum expressions for
$I_2(u),I_3(u)$ commute but it is not true in general for the higher integrals.

\paragraph{Main result. }
Consider the quantum Lax operator given by the expression
(\ref{Lax}) where $\Phi_i$ are matrices with elements in
$U(\g)^{\otimes k}$ such that ${(\Phi_i)}_{kl}$ is the $e_{kl}$ generator
of $\g$ lying in
the $i$-component of the tensor product.
Consider the following differential operators applied to the function $1$:
\begin{equation}
\label{qi}
QI_i(u)=Tr_{1,\ldots,n}A_n (L_1(u)-\partial_u)(L_2(u)-\partial_u)\ldots
(L_i(u)-\partial_u) \mathbf{1}
\end{equation}
where $A_k$ is the antisymmetrizer in $Mat_{n\times n}^{\otimes n}.$
\\
{\bf Main Theorem. \it The quantities $QI_i(u)$ constitute a
commutative family in $U(\g)^{\otimes k}(u)$ in the following sense
$$[QI_i(u),QI_j(v)]=0,\qquad\forall i,j=1,\ldots n,~\forall u,v;$$
and their classical limit in the sense of associated graded algebra
coincides with the family of classical Gaudin hamiltonians.}

\paragraph{Plan of the paper.}
In section \ref{cf} we recall the construction of the Yangian $Y(\g)$ and
the Bethe subalgebra in it, which is a maximal commutative subalgebra for
general choice of parameters. Further, considering the
$k$-th tensor power of the evaluation
homomorphism we obtain a commutative
family in  $U(\g)^{\otimes k}$. In section \ref{cl} we extend the Bethe subalgebra to
the commutative family of formal expressions in $\h.$ This formal family
gives the standard Bethe subalgebra at $\h=1,$ whereas we use its special
limit at $\h=0.$ We also slightly change a basis using only linear
transformations.
The main result of the paper is contained in section
\ref{ar}.

\paragraph{Acknowledgement.} The author would like to thank his colleagues
in ITEP for numerous discussions and stimulating atmosphere, especially A.
Chervov, L. Rybnikov, as well as S. M. Kharchev and S. M. Khoroshkin for useful remarks.

\section{Quantum Gaudin system}
\subsection{Bethe subalgebra} \label{cf}
The Yangian was introduced by Drinfeld \cite{Dr} and take an important role
in describing rational solutions of the Yang-Baxter Equation.  $Y(\mathfrak{gl}_n)$ is generated by
elements $t_{ij}^{(k)}$ subject to specific relations on the generating function
$T(u)$ with values in $Y(\mathfrak{gl}_n)\otimes End(\mathbb{C}^n)$
$$T(u)=\sum_{i,j} E_{ij}\otimes t_{ij}(u),\qquad
t_{ij}(u)=\delta_{ij}+\sum_k t_{ij}^{(k)} u^{-k},$$ where $E_{ij}$ are matrices
with $1$ on the $i,j$-th place. The relations on this generating function
involve the Yang $R$-matrix $$R(u)=1-\frac 1 u \sum_{i,j}E_{ij}\otimes
E_{ji}$$ and reads as the following equation
$$R(z-u)T_1(z)T_2(u)=T_2(u)T_1(z)R(z-u)$$ in $End(\mathbb{C}^n)^{\otimes
2}\otimes Y(\mathfrak{gl}_n)[z,u],$
where $$T_1(z)=\sum_{i,j}E_{ij}\otimes {1}\otimes t_{ij}(z),\quad
T_2(u)=\sum_{i,j} {1}\otimes E_{ij}\otimes t_{ij}(u).$$
There is a known realization of $Y(\mathfrak{gl}_n)$ in $U(\mathfrak{gl}_n)$ i.e.
a homomorphism $\rho_1:Y(\mathfrak{gl}_n)\rightarrow U(\mathfrak{gl}_n)$
\begin{equation}\label{gl}
T(u)=1+\frac 1 u \sum_{i,j}E_{ij}\otimes e_{ij}\stackrel{def}{=}:1+
\frac \Phi u ,
\end{equation}
where $e_{ij}$ are
generators of $\g.$
The following expression also provide a realization of Yangian in
$U(\mathfrak{gl}_n)^{\otimes k}$ for the given $k$-tuple
of complex numbers $\alpha=(z_1,\ldots,z_k)$
$$T^{\alpha}(u)=T^1(u-z_1)T^2(u-z_2)\dots T^k(u-z_k),$$ where $T^l(u-z_l)$ is
the realization given by (\ref{gl}) with $e_{ij}$ lying in the $l$-th component of
$U(\mathfrak{gl}_n)^{\otimes k}.$ Let $\rho_{\alpha}$ be the corresponding
mapping $\rho_\alpha:Y(\mathfrak{gl}_n)\rightarrow U(\mathfrak{gl}_n)^{\otimes k}$.
{\Lem \label{YB} $\rho_\alpha$ is a homomorphism. This is equivalent to the
relation
$$R(z-u)T_1^{\alpha}(z)T_2^{\alpha}(u)=T_2^{\alpha}(u)T_1^{\alpha}(z)R(z-u)$$
in $End(\mathbb{C}^n)^{\otimes
2}\otimes U(\g)^{\otimes k}[z,u].$}
\\
{\bf Proof. } Here we use the fact that $T_1^l(z)$ commutes with $T_2^k(u)$
if $l\ne k.$ Indeed,
$$ R(z-u)T_1^1(z-z_1)T_1^2(z-z_2)\dots T_1^k(z-z_k)
T_2^1(u-z_1)T_2^2(u-z_2)\dots T_2^k(u-z_k)=$$
$$ R(z-u)T_1^1(z-z_1)T_2^1(u-z_1)T_1^2(z-z_2)\dots T_1^k(z-z_k)
T_2^2(u-z_2)\dots T_2^k(u-z_k)=$$
$$T_2^1(u-z_1) T_1^1(z-z_1) R(z-u)T_1^2(z-z_2)\dots T_1^k(z-z_k)
T_2^2(u-z_2)\dots T_2^k(u-z_k)=$$
et cetera...
\\
$\square$

The simple way to obtain the commuting family in
$U(\mathfrak{gl}_n)^{\otimes k}$ is to take the image
$\rho_{\alpha}(Z(Y(\mathfrak{gl}_n)))$ of the center of Yangian. The
homomorphism $\rho_\alpha$ is not surjective and that is why this commuting
family is not trivial (contains not only constants). The center of Yangian
is given by $$qdet T^{\alpha}(u)=\sum_{\sigma\in S_n} sgn(\sigma)
t^{\alpha}_{\sigma(1),1}(u)\dots t^{\alpha}_{\sigma(n),n}(u-n+1)$$ see for
example Proposition $2.7$ from \cite{Molev}. Unfortunately this family is
too small, $T^{\alpha}(u)$ is a rational function on $u$ with $k$ simple
poles, $qdet T^{\alpha}(u)$ has $kn$ simple poles, the corresponding residues
play the role of formal quantum hamiltonians whereas the dimension of the
classical phase space is of order $kn^2.$ There is another
construction enlarging the family of central elements in Yangian up to a
maximal commutative subalgebra (see \cite{KR},\cite{NO} or $2.14$ of \cite{Molev1}):
let $C$ be an $n\times n$ matrix, $T(u)$ the generating function for the
elements of $Y(\g)$, $A_n$ is the matrix of antisymmetrizer in
${\mathbb{C}^n}^{\otimes n}$ and $$T_i(u) = \sum_{kl} 1\otimes  \ldots
\otimes 1\otimes
 \stackrel{i}{E_{kl}}\otimes 1 \otimes \ldots\otimes 1 \otimes t_{kl}(u)$$
 then the coefficients of
$$\tau_i(u)=Tr A_n T_1(u)T_2(u-1)\dots T_i(u-i+1)C_{i+1}\dots C_n
\quad i=1,\ldots n $$
generate the commuting family
$$[\tau_i(u),\tau_j(v)]=0$$
which is maximal if the matrix $C$ has simple
spectra. In the realization $T^{\alpha}(u)$ each $\tau_i(u)$ is a rational
function on $u$ with $ik$ simple poles, the total number of formal quantum
conserved quantities is equal to $k\frac {(n+1)n} 2$ which is approximatively
half the dimension of the classical phase space.

\subsection{Formal deformation} \label{cl}
Let us consider the canonical rescaling of the universal enveloping algebra
obtained by multiplying structure constants by $\h.$ The corresponding
$\g$ representation of Yangian $T(u)$ looks as follows $$T(u)=1+\frac
{\h \Phi} u.$$ It solves YBE with the rescaled Yang R-matrix
$$R(u)=1-\frac {\h P} u.$$ The construction of the commuting family
is reproduced literally in this case:
\begin{equation}\label{tau}
\boxed{\tau_i(u)=(Tr A_n T^{\alpha}_1(u)T^\alpha_2(u-\h)\dots
T^\alpha_i(u-(i-1)\h)C_{i+1}\dots C_n)}
\quad i=1,\ldots n .
\end{equation}
We will use later the following obvious
{\lem \label{first} Let two formal in $\h$ expressions
$$M(\h)=\sum_{i=m}^\infty M_i \h^i\quad\mbox{and}\quad
N(\h)=\sum_{i=n}^{\infty}N_i \h^i$$
with values in some associative algebra $\mathcal{A}$ commute, then their
first coefficients also commute
$$[M_m,N_n]=0.$$
It is also true for the first non-central coefficients of formal series.
}

{\Rem Let us note that the classical Gaudin Lax operator can be recovered
as a classical limit of $T^\alpha:$
$$T^\alpha(u)=1+\h L(u)+O(\h^2)$$ where
$$L(u)=\sum_i\frac {\Phi_i} {u-z_i}$$ is the first non-central coefficient of
the operator-valued matrix. This expression lies in the first filtration
component and after taking classical limit remains of the same form (with described
in Introduction exchange of the meaning of matrix elements).}

\subsection{Higher Hamiltonians} \label{ar}
Let us introduce ``quasiclassic'' hamiltonians:
$$\boxed{I_k=Tr A_n L_1(u) L_2(u)\ldots L_k(u).}$$
Their classical limits in the sense of ``Gr'' are given by the following
expressions:
\begin{equation}
\begin{split}
i_1&=Tr L(u) (n-1)!;\\
i_2&=(Tr^2L(u)-TrL^2(u))(n-2)!;\\
i_3&=(2TrL^3(u)-3TrL(u)Tr L^2(u)+Tr^3L(u))(n-3)!
\end{split}
\end{equation}
and can be taken as a basis of classical Gaudin hamiltonians.

Consider formal expressions
\begin{equation}\label{sk}
\boxed{s_i=\sum_{j=0}^{i}(-1)^j C_i^j\t_{i-j}}
\end{equation}
where $\tau_0=Tr A_n 1=n!.$
{\lem $s_i$ has the following expansion on $\h$
$$s_i(u)=\h^i QI_i(u)+O(\h^{i+1})$$
where $QI_i$ is given by the formula
\bea \label{qi1}
\boxed{QI_i(u)=Tr_{1,\ldots,n}A_n (L_1(u)-\partial_u)(L_2(u)-\partial_u)\ldots
(L_i(u)-\partial_u) \mathbf{1}}
\eea
where $A_i$ is the antisymmetrizer matrix in $Mat_{n\times n}^{\otimes i}$
and $L_j(u)$ is the quantum Lax operator (\ref{Lax}) lying in the $j$-th
tensor component in $Mat_{n\times n}^{\otimes i}.$
}
\\
{\bf Proof.~ }
Let us represent the formula for the generators of the Bethe subalgebra (\ref{tau})
in a quite different manner:
\begin{equation}
\label{tau1}
\tau_i(u-\h)e^{\d -i\h\p}=Tr A_n e^{\d-\h\p}T_1(u)e^{\d-\h\p}T_2(u)\ldots
e^{\d-\h\p}T_i(u)
\end{equation}
where expressions on both sides are differential operators on $u$ with
values in $U(\g)^{\otimes k}(u).$
These formulas can be combined in a sort of a generating function:
\bea
&& Tr A_n
(e^{\d-\h\p}T_1(u)-1)(e^{\d-\h\p}T_2(u)-1)\ldots(e^{\d-\h\p}T_i(u)-1)\nn\\
&=&\sum_{j=0}^i\tau_j(u-\h)(-1)^{i-j}C_i^j e^{\d(j-i)\h\p} \label{cr}
\eea
Applying then both sides to the function $1$ and using the fact that
$$e^{\d-\h\p}T(u)-1=\h(L(u)-\p)+O(\h^2)$$
we obtain
\bea
s_i(u)=\h^iTr A_n (L_1(u)-\partial_u)(L_2(u)-\partial_u)\ldots
(L_i(u)-\partial_u) \mathbf{1}+O(\h^{i+1}).\nn
\eea
$\square$
\\
Hence we have obtained that $QI_i$ commute. Indeed, $[s_i(u),s_j(v)]=0$ due
to the fact that $s_i(u)$ are linear combinations of generators of the Bethe
subalgebra. Then, $[QI_i(u),QI_j(v)]=0$ in virtue of lemma \ref{first}.
It is quite evident that the classical limit of $QI_i$ coincides with the
classic hamiltonian, the element of the highest order in (\ref{qi1}) is
$$Tr A_n L_1(u)L_2(u) \ldots L_i(u).$$
\\
Thus we proved the main theorem of this paper:
{\Th The quantities $QI_i(u)$ provide a  quantization of the
classical Gaudin model.}
\vskip 1cm
Let us analyze the obtained quantum hamiltonians at small order:
\begin{itemize}
  \item {$\mathbf l=1$~}
$$s_1=\h TrL(u)(n-1)!+O(\h^2);$$
the first term at $\h$ is just the ``semiclassic'' hamiltonian $I_1(u).$
  \item {$\mathbf l=2$~}
$$s_2=\t_2-2\t_1+\t_0=\h^2(-\partial_u Tr L(u)(n-1)!+TrA_nL_1(u)L_2(u))+O(\h^3);$$
the first term at $\h^2$ can be expressed as $$I_2(u)-\partial_u I_1(u).$$
  \item {$\mathbf l=3$~}
\bea
s_3&=&\h^3Tr A_n(\partial^2_u
L_1(u)-\partial_uL_1(u)L_2(u)-2L_1(u)\partial_uL_2(u)\nn\\
&+&L_1(u)L_2(u)L_3(u))+O(\h^4).
\nn
\eea
The first term at $\h^3$ can be simplified as follows:
$$I_3(u)-\frac 32\partial_u I_2(u)+\partial^2_u I_1(u);$$ where we have used
the property $Tr A_n \partial_u L_1 (u) L_2(u)=Tr A_n L_1(u)\partial_u
L_2(u)$ which is not true for the  number of multipliers greater then $2$.
\item {$\mathbf l=4$~}
\bea
s_4&=&\h^4Tr A_n(-\partial^3_u L_1(u)+\partial^2_u L_1(u) L_2(u)+
3L_1(u)\partial^2_u L_2(u) + 3 \partial_u L_1(u)\partial_u L_2(u)\nn\\
&-&\partial_uL_1(u)L_2(u)L_3(u)-2L_1(u)\partial_uL_2(u)L_3(u)
-3L_1(u)L_2(u)\partial_uL_3(u)\nn\\
&+&L_1(u)L_2(u)L_3(u))L_4(u)) +O(\h^5).
\nn
\eea
The first term here at $\h^4$ also can be simplified:
\bea
&&I_4(u)-\partial_u I_3(u)+2\partial^2_u I_2(u)-\partial^3_u
I_1(u)\nn \\
&-&Tr A_n(\partial_u L_1(u)\partial_u L_2(u)+
L_1(u)\partial_uL_2(u)L_3(u)+2L_1(u)L_2(u)\partial_uL_3(u))
\nn
\eea
\end{itemize}
The arguments of the previous section give us that the first coefficients as
like as their derivatives commute and we obtain that
the quasiclassic hamiltonians for
$i=1,\ldots,3$ lie in our commutative family.
However the higher hamiltonians $i\ge 4$ have unavoidable
quantum corrections.

\subsection*{Concluding remarks}
\paragraph{Pull back to $Y(\g)$}
The proposed construction for quantum hamiltonians can be realized in a
more general setup, it can be pulled back to the Yangian. In this case it
provides a commutative family expressed only in the first order generators
of the Yangian
$$QI_i(u)=Tr_{1,\ldots,n}A_n (T_1^{(1)}(u)-\partial_u)(T^{(1)}_2(u)-\partial_u)\ldots
(T^{(1)}_i(u)-\partial_u) \mathbf{1}.$$
In this way one could expect to quantize a quite
wider class of models: rational matrices with higher order poles and the
Hitchin system on singular curves.
\paragraph{Bethe ansatz}
The obtained formulas for quantum hamiltonians provide a kind of
decomposition which could be tractable in the context of the Bethe ansatz
for higher dimensional spin components of the Gaudin model.

\end{document}